# Engaging Users with Educational Games: The Case of Phishing

**Matt Dixon**
Northumbria University
Newcastle, UK
m.dixon@northumbria.ac.uk

**Nalin Asanka Gamagedara Arachchilage**
University of New South Wales
Australia
nalin.asanka@adfa.edu.au

**James Nicholson**
Northumbria University
Newcastle, UK
james.nicholson@northumbria.ac.uk

## ABSTRACT

Phishing continues to be a difficult problem for individuals and organisations. Educational games and simulations have been increasingly acknowledged as versatile and powerful teaching tools, yet little work has examined how to engage users with these games. We explore this problem by conducting workshops with 9 younger adults and reporting on their expectations for cybersecurity educational games. We find a disconnect between casual and serious gamers, where casual gamers prefer simple games incorporating humour while serious gamers demand a congruent narrative or storyline. Importantly, both demographics agree that educational games should prioritise gameplay over information provision – i.e. the game should be a game with educational content. We discuss the implications for educational games developers.

## 1 INTRODUCTION

Cybersecurity is a growing issue for both individuals and organisations as more and more facets of life are integrated into the internet and digital spaces. Any given individual or organisation can have a wealth of sensitive personal or financial information stored online and on their computer(s) and regularly input this information online to access and use services such as online banking and shopping [1, 2]. As such, phishing proves to be a highly profitable method of utilising social engineering tactics to manipulate individuals into disclosing valuable information or unwittingly providing access to their device or accounts [14]. Phishing acts as a low risk high reward scamming method to gain a variety of critically valuable pieces of information – from credit card numbers to highly-sensitive political emails (e.g. John Podesta in 2016). Successful phishing attempts accounted for 75% of security breaches in 2017 [8] with an estimated cost of $3.5 million per breach in large organisations [6]. Evidently, phishing represents a widespread and devastating threat for modern internet users.

One method for dealing with social engineering attacks such as phishing is by exposing employees to better organisational training tailored to specific and relevant threats (e.g. [10]). However, employee engagement with training has been historically poor. Educational games and simulations, on the other hand, have become increasingly acknowledged as an enormous and powerful teaching tool that may result in an "instructional revolution" [2, 7, 13]. The main reason is that game-based education allows users to learn through experience and the use of virtual environment while leading them to approach problem solving through critical thinking [5]. In addition, game-based education is further useful in motivating players to change their behaviour [2]. Researchers have revealed that well designed end-user education is imperative to combat against phishing attacks [2, 13]. Furthermore, game-based learning can be effective not only for changing people's behaviour, but also for developing their logical thinking to solve mathematical problems. Arachchilage et al. [2] designed and developed a mobile game that aimed to enhance the users' avoidance behaviour through motivation to protect themselves against phishing threats.

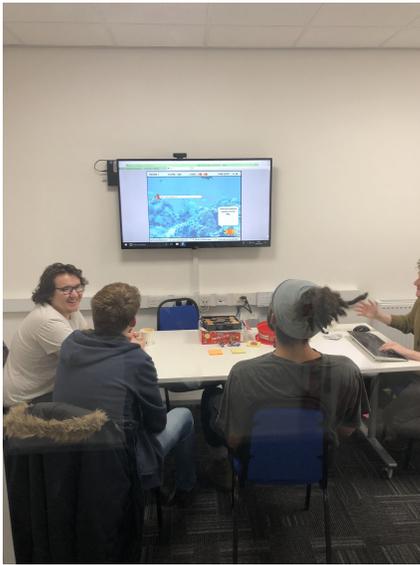

**Figure 1: Participants playing Anti-Phishing Phil in a workshop session**

However, there has been a lack of work on investigating how one can design a gamified approach that users can better engage in order for them to interact with the game, which will eventually contribute to enhance their learning experience [2, 13].

Therefore, in the current study, we set out to engage young internet users in focus groups to understand their attitudes towards educational games. In this paper, we report on a subset of findings related to general game design, and on the cybersecurity topic of phishing.

## 2 METHOD
### Design
This study explored 9 young adults' (mean age: 22 years old; 2 female) attitudes towards educational games – more specifically their perceptions of engagement. We recruited younger adults aged 18-25 with a range of gaming habits to capture a demographic that engages both with games and online activities, but that is also targeted by social engineering attacks [9] and that has been shown to exhibit security vulnerabilities in the past [12]. Three 80-minute workshops, consisting of three participants each, were carried out by a member of the research team and were audio recorded. After transcription, the lead author proceeded to code the data, and developed themes using the thematic analysis framework identified by Braun and Clarke [3]. The second and third authors then inspected the codes and themes for consistency and relevancy in the process of researcher triangulation.

### 3.2.1 Story
The main character of the game is Phil, a young fish living in the Interweb Bay. Phil wants to eat worms so he can grow up to be a big fish, but has to be careful of phishers that try to trick him with fake worms (representing phishing attacks). Each worm is associated with a URL, and Phil's job is to eat all the real worms (which have URLs of legitimate web sites) and reject all the bait (which have phishing URLs) before running out of time. The other character is Phil's father, who is an experienced fish in the sea. He occasionally helps Phil out by giving Phil some tips on how to identify bad worms (and hence, phishing web sites).

**Figure 2: Game story of the Anti-Phishing Phil**

### Procedure
Participants were invited into the lab in groups of three for a workshop lasting approximately 80 minutes. Participants were first introduced to the concept of phishing, and were then asked about their attitudes towards learning from games. Next, participants were given the chance to play a round of Anti-Phishing Phil [13], first demonstrated by the facilitator and then played in turn by each participant (see Figure 4. We chose Anti-Phishing Phil as it is a well-known and used game used both in academia and industry, as well as an inspiration for other modern cybersecurity games [2, 4, 11]. They were then asked to comment on the game and provide any feedback related to improvements of mechanics and theme. Participants were also encouraged to discuss their ideas for designing an educational phishing game while being provided with possible content (e.g. facts and procedures), and were afforded sticky notes, various sized sheets of paper, and pens to illustrate desired interactions and stories. Finally, participants were once again asked about their attitudes towards gaming and learning from games.

## 3 RESULTS
### Regional Differences
Participants were asked various questions and given the chance to discuss a range of topics and ideas with the rest of the group across the workshop. After playing Anti-Phishing Phil (APP) [13], participants were asked to give their opinion on the game as a learning aid. Initial opinions were mixed but a general consensus revolved around APP being a suitable game to teach children the basics of phishing. There were a number of issues and criticisms of APP, one of which was the cultural bias within the game – namely that many of the websites in question were for American companies and thus unfamiliar to those outside of the United States. As a result, some participants found it difficult to assess whether aspects of the URL were suspicious or part of a company name (e.g. "CitiBank" where participants were unsure whether *Citi* was part of the name or a misspelling).

*3.2.2 Mechanics*
The game is split into four rounds, each of which is two minutes long. In each round, Phil is presented with eight worms, each of which carries a URL that is shown when Phil moves near it (see Figure 1). The player uses a mouse to move Phil around the screen. The player uses designated keys to "eat" the real worms and "reject" the bait. Phil is rewarded with 100 points if he correctly eats a good worm or correctly rejects a bad one. He is slightly penalized for rejecting a good worm (false positive) by losing 10 seconds off the clock for that round. He is severely penalized if he eats a bad worm and is caught by phishers (false negative), losing one of his three lives. We developed this scoring scheme to match the real-world consequences of falling for phishing attacks, in that correctly identifying real and fake web sites is the best outcome, a false positive the second best, and a false negative the worst. The consequences of Phil's actions are summarized in Table 2.

**Figure 3: Game mechanics of the Anti-Phishing Phil [13]**

*3.2.3 Technology*
The game is implemented in Flash 8. The content for the game, including URLs and training messages, are loaded from a separate data file at the start of the game. This provides us with a great deal of flexibility and makes it easy to quickly update the content. In each round of the game, four good worms and four phishing worms are randomly selected from the twenty URLs in the data file for that round. We also use sound and graphics to engage the user better. This includes sound effects to provide feedback on actions, background music, and underwater background scenes.

**Figure 4: Game technology of the Anti-Phishing Phil [13]**

### Casual and Serious Gamers

We noticed a clear divide between casual and serious gamers when it came to the preferred design of educational games. Casual gamers suggested that educational games should utilise humour and consist of an overall light-hearted theme, e.g. *"you could make it not to be taken seriously. I feel like the better way to get through to adults these days is like inventing games and taking the mick"*. Two participants acknowledged that they engaged with APP as they found the theme silly and humourous: *"well we found it funny and as a result we have in turn actually paid attention as a result of it"*.

On the other hand, serious gamers believed that an educational game should be immersive and present a strong narrative to engage players in the learning process, for example by mimicking themes and concepts from existing AAA games: *"I think thematically you could have a detective has received all these ransom notes and they've all got URLs for them and he's got to pick the one that's not gonna get him hacked. So think L.A Noire but with phishing."*. Other serious gamers, on the other hand, believed that simply having an engaging story-line would be beneficial, without the need for expensive mechanics or graphics: *"yeah a TellTale game would work... some sort of reason to keep you going with it, like if there's a company conspiracy, I don't know just be fun, like if there's some sort of conspiracy for you to uncover and you'd have to learn these tools to actually be able to get through the story"*.

### Games for Teaching and Games for Learning

An interesting theme that arose throughout the workshops was the idea of covert learning as an important factor in engaging adults who may be less open to using a game to learn. Participants believed that most adults would not engage in a game like APP because of the childish theme and overt teaching style – in fact, participants suggested that a game should stand as an enjoyable experience by itself and not be developed solely as an educational tool. By smoothly integrating information about phishing as part of the game's narrative, people could be less averse to playing an educational cyber-security game and learn the required material by picking it up as part of a fun game: *"so not consciously learning, you're just picking up the information – it's becoming engrained"*. Interestingly, participants across workshops agreed on the sentiment that to engage users with educational games, developers had to *"either take the mick or you get them there subconsciously"*. Specifically, Participant S – not a keen gamer – claims that if she were to pick up a game, it would be to have fun and not something too serious such as a game based in a simulation of an email/browser. Participant B - a more dedicated gamer - shares the sentiment, claiming that *"when you're making games that are educational, people tend to make it too educational and not gamey and then it's lost"*. As such, the information provided to participants should be naturally engrained into the content of the game, a far cry from the APP fusion of fishing and cybersecurity.

## 4 DISCUSSION

After discussing the many facets of phishing, viewing practical examples of phishing tactics in action, and demoing an existing cybersecurity educational game, participants across the focus groups provided a range of interesting and unique thoughts and attitudes towards the development of educational games. One of the most striking and consistent themes was the concept of overt and covert learning within a game. Participants consistently believed that when using a video game as a learning medium it should primarily be just that – a video game. The learning aspect should be naturally integrated into the game, so the player picks up information and skills through playing the game, e.g. *"I feel games and education can be mixed a lot better"* In this sense, the theme and learning outcomes of the game should be congruent to create a smooth experience in which the player is taught and tested through playing the game without necessarily realising that learning is taking place. This contrasts with existing education games (e.g. APP) where the goal of the game is clearly to teach players phishing concepts, with very little narrative or theme development.

Engagement and enjoyment of the game was very important across both casual and more devoted gamers. Casual gamers, those who only pick up a game occasionally for fun, expect a smooth, enjoyable and simple experience. They likely would not seek out or engage with a game designed for learning if it does not integrate the knowledge it seeks to teach into a fun, engaging experience. If the game offers knowledge and gameplay as two distinct features, then from the perspective of the casual gamer, they may not see any reason to use such a game, as the entertainment value of the game will generally pale compared to mainstream games and the educational content could simply be found in other traditional formats such as articles or videos. More devoted gamers will typically be more in tune with the harmony of the content, story and mechanics, therefore if the knowledge is presented overtly and separately from the gameplay, then we have a similar problem to casual gamers.

These findings have important implications for game developers. Firstly, they highlight the challenge that they face when designing educational games, and the need to cater for both casual and serious gamers. While this is not a problem associated solely with educational games, it is important to consider the context in which they will be deployed – e.g. the workplace is likely to be an important market. Not much is known about the composition of workplaces with regards to their gaming preferences, and future work could explore this in more depth in order to understand what types of educational games are more likely to succeed in these environments.

Secondly, the incorporation of gameplay, themes, and educational content appears to be an extremely challenging but important aspect for the design of educational games. Specifically, most participants expressed how an effective educational game would consist of an engaging game with covert learning, rather than the more traditional overt learning that appear to turn off potential players. However, more research is needed in order to understand what other factors play a role in users' engagement with educational cybersecurity games, and how best to incorporate all these factors into a tool that is both engaging but also effective at knowledge transmission.

## REFERENCES

[1] Nalin Asanka Gamagedara Arachchilage and Steve Love. 2014. Security awareness of computer users: A phishing threat avoidance perspective. *Computers in Human Behavior* 38 (2014), 304–312.
[2] Nalin Asanka Gamagedara Arachchilage, Steve Love, and Konstantin Beznosov. 2016. Phishing threat avoidance behaviour: An empirical investigation. *Computers in Human Behavior* 60 (2016), 185–197.
[3] Virginia Braun, Victoria Clarke, and Gareth Terry. 2014. Thematic analysis. *Qual Res Clin Health Psychol* 24 (2014), 95–114.
[4] Gamze Canova, Melanie Volkamer, Clemens Bergmann, and Benjamin Reinheimer. 2015. NoPhish app evaluation: lab and retention study. *USEC. Internet Society* (2015).
[5] Ching-Yi Chang and Gwo-Jen Hwang. 2019. Trends in digital game-based learning in the mobile era: a systematic review of journal publications from 2007 to 2016. *International Journal of Mobile Learning and Organisation* 13, 1 (2019), 68–90.
[6] Ponemon Institute. 2018. *2018 Cost of a Data Breach Study: Global Overview*. Technical Report. IBM.
[7] Jurjen Jansen and Paul van Schaik. 2019. The design and evaluation of a theory-based intervention to promote security behaviour against phishing. *International Journal of Human-Computer Studies* 123 (2019), 40–55.
[8] Rebecca Klahr, Sophie Amili, Jayesh Navin Shah, Mark Button, and Victoria Wang. 2016. Cyber security breaches survey 2016. *UK Government, Ipsos MORI and University of Portsmouth. DOI= http://bit.ly/1T4MveX* (2016).
[9] Aamna Mohdin. 2018. Scammers target students with fake tax refund emails. https://www.theguardian.com/money/2018/nov/17/scammers-target-students-with-fake-tax-refund-emails
[10] James Nicholson, Lynne Coventry, and Pam Briggs. 2018. Introducing the cybersurvival task: assessing and addressing staff beliefs about effective cyber protection. In *Fourteenth Symposium on Usable Privacy and Security (SOUPS 2018)*. 443–457.
[11] Sankalp Pandit, Sukanya Vaddepalli, Harshal Tupsamudre, Vijayanand Banahatti, and Sachin Lodha. 2018. PHISHY-A Serious Game to Train Enterprise Users on Phishing Awareness. In *Proceedings of the 2018 Annual Symposium on Computer-Human Interaction in Play Companion Extended Abstracts*. ACM, 169–181.
[12] Steve Sheng, Mandy Holbrook, Ponnurangam Kumaraguru, Lorrie Faith Cranor, and Julie Downs. 2010. Who falls for phish?: a demographic analysis of phishing susceptibility and effectiveness of interventions. In *Proceedings of the SIGCHI Conference on Human Factors in Computing Systems*. ACM, 373–382.
[13] Steve Sheng, Bryant Magnien, Ponnurangam Kumaraguru, Alessandro Acquisti, Lorrie Faith Cranor, Jason Hong, and Elizabeth Nunge. 2007. Anti-phishing phil: the design and evaluation of a game that teaches people not to fall for phish. In *Proceedings of the 3rd symposium on Usable privacy and security*. ACM, 88–99.
[14] Amber A Smith-Ditizio and Alan D Smith. 2019. Computer Fraud Challenges and Its Legal Implications. In *Advanced Methodologies and Technologies in System Security, Information Privacy, and Forensics*. IGI Global, 152–165.